\documentstyle[12pt]{article}
\newlength{\digitwidth} 
\catcode`?=\active \def?{\kern\digitwidth}

\newcommand{\be}{\begin{equation}}
\newcommand{\ee}{\end{equation}}
\newcommand{\bea}{\begin{eqnarray}}
\newcommand{\eea}{\end{eqnarray}}
\newcommand{\brr}{\begin{array}}

\newcommand{\tr}{{\rm Tr}}
\newcommand{\op}{\hat O^{\Delta S =2} }

\newcommand{\MSbar}{\overline{\mbox{\small MS}}}

\begin{document}
\pagestyle{empty}
\begin{flushright}
BU-HEP 95-26 \\
CERN-TH/95-234  \\
ROME prep. 1105/95 \\
\end{flushright}
\centerline{\LARGE{\bf{Chiral  behaviour of the  lattice $B_K$-parameter}}}
\centerline{\LARGE{\bf{with the Wilson and Clover Actions at $\beta = 6.0$}}}
\vskip 0.3cm
\centerline{\bf{M. CRISAFULLI$^a$, A. DONINI$^{a,b}$, V. LUBICZ$^c$,
G. MARTINELLI$^{b,*}$, }}
\centerline{\bf{ F. RAPUANO$^{a}$, M. TALEVI$^a$, C. UNGARELLI$^d$,
A. VLADIKAS$^{e}$}}
\centerline{$^a$ Dip. di Fisica, Univ. di Roma ``La Sapienza'' and
INFN, Sezione di Roma,}
\centerline{P.le A. Moro, I-00185 Rome, Italy.}
\centerline{$^b$ Theory Division, CERN, 1211 Geneva 23, Switzerland.}
\centerline{$^c$ Dept. of Physics, Boston University, Boston MA 02215,
USA.}
\centerline{$^d$ Dip. di Fisica, Univ. di Pisa and INFN, Sezione di Pisa}
\centerline{P.za Torricelli 2, I-56100 Pisa, Italy.}
\centerline{$^e$ INFN,
Sezione di Roma II and Dip. di Fisica, Univ. di Roma ``Tor Vergata'',}
\centerline{Via della Ricerca Scientifica 1, I-00133 Rome, Italy.}
\date{}
\abstract{
\par\noindent
We present results for the kaon $B$-parameter $B_K$ from a sample
of $200$ configurations using the Wilson action and $460$ configurations
using the SW-Clover action, on a $18^3 \times 64$ lattice at $\beta=6.0$.
We compare results obtained by renormalizing the relevant operator
with  different ``boosted" values of the strong coupling constant
 $\alpha_s$. In the case of the SW-Clover action, we also use the
operator renormalized  non-perturbatively. In the Wilson case,
we observe a strong dependence of $B_K$ on the prescription
adopted for $\alpha_s$, contrary to the results of the Clover case which are
almost unaffected by the choice of the coupling. We also find that
the matrix element of the  operator renormalized non-perturbatively
has a better chiral behaviour. This gives us our  best estimate
of the renormalization group invariant $B$-parameter, $\hat
B_K=0.86 \pm 0.15$.}
\vskip 0.8cm
\begin{flushleft}
CERN-TH/95-234 \\
September 1995
\end{flushleft}
\vskip 0.5 cm
\centerline{$^*$ On leave of absence from Dip. di Fisica,
Universit\`a degli Studi ``La Sapienza'', Rome, Italy.}
\vfill\eject
\pagestyle{empty}\clearpage
\setcounter{page}{1}
\pagestyle{plain}
\newpage
\pagestyle{plain} \setcounter{page}{1}

\section{Introduction}
\label{sec:introduction}

The kaon $B$-parameter $B_K$ is a quantity of great phenomenological interest,
being related to the $\epsilon$-parameter which measures $CP$-violation
in the $K^0$--$\bar K^0$ system.  With a large top quark mass
\cite{TOP}, an accurate prediction of $B_K$ enables us to limit
 the range of values of the CP-violation phase $\delta$;
see for example ref. \cite{CIUCHINI2}.  $B_K$ is  obtained
from the
calculation of the matrix element of the relevant four-fermion
operator
 $\langle \bar K^0 \vert {\hat O}^{\Delta S = 2} \vert K^0 \rangle$,
where ${\hat O}^{\Delta S = 2}={\bar s} \gamma^L_{\mu} d {\bar s}
\gamma^L_{\mu}  d$;  $s$ and $d$ stand for strange and down
quarks and $\gamma^L_{\mu}=\frac{1}{2} \gamma_{\mu} ( 1 - \gamma_5 )$.
A non-perturbative estimate of the matrix element may be obtained in the
framework of lattice QCD \cite{CMP,BGGM}.
Several lattice calculations
of $B_K$ have been performed in recent years, both with staggered
\cite{SHARPE,jap} and Wilson \cite{GAVELA}--\cite{GUPTAB} fermions.
\par
With the Wilson or SW-Clover quark lattice actions,
 due to the presence of the  chiral symmetry breaking  term,
 $\hat O^{\Delta S=2}$ mixes with operators of different chirality
\cite{BOCHICCHIO,MARTIW}. For this reason,
it is possible to define a renormalized operator with definite chiral
properties only in the continuum limit, i.e. when $a \to 0$. At finite $a$,
one can improve the chiral behaviour of the matrix element
of $\hat O^{\Delta S=2}$  by sub\-tracting
a suitable set of dimension-6 operators. The mixing coefficients
have been computed so far only in one-loop perturbation theory
\cite{MARTIW}--\cite{BORRELLI}. In
this way, the systematic error in the value of the matrix element
determined on the lattice  is of $O(\alpha_s^2)$ and, due to the finitess
of the lattice spacing, of
$O(a)$.  Following the Symanzik proposal, one can reduce
the discretization errors  from $O(a)$ to $O(\alpha_s a)$ by using the
tree-level ``improved''  SW-Clover  quark action \cite{SW,HEATLIE}.
 Using this action, the
improvement has been shown to be  effective for two-fermion operators,
 at  values of $\beta$ currently used
in numerical simulations \cite{ZETA_V}--\cite{WIUKQCD}.
It remains true, however, that ignorance of
higher-order perturbative corrections to the mixing coefficients, which are
expected to be large  \cite{LEPAGE}, can distort the chiral behaviour
of the operator and hence induce a large
systematic error in the determination of  $B_K$.  For this reason
the  kaon matrix element of ${\hat O}^{\Delta S = 2}$ does not
vanish in the chiral limit \cite{BERNARD1}.
This renders the evaluation of $B_K$ more problematic with Wilson
or SW-Clover  fermions than with staggered fermions.

In this paper, we present an extended study of
$\langle {\bar K}^0 \vert {\hat O}^{\Delta S = 2} \vert K^0 \rangle $,
using different sets of quenched gauge field configurations,  at $\beta=6.0$.
We address the two  systematic effects which afflict its chiral behaviour:
corrections of $O(\alpha_s^2)$ or higher in the mixing coefficients and
contributions of $O(a)$ due to the finiteness of the lattice spacing $a$.
 In order to estimate both sources of error, we have performed a high
statistics analysis of these effects.  Discretization errors
have been studied by comparing the results for $B_K$  obtained with the
Wilson and SW-Clover actions. The systematic
error due to the use of perturbation theory
in  the calculation of the mixing coefficients
 has also been analysed by studying the
effects of a different choice of the expansion  parameter, as proposed
in ref. \cite{LEPAGE}. We also used, in the Clover case,
 the coefficients computed in \cite{TALEVI} with  the
 Non-Perturbative Method (NPM) of ref. \cite{NP}.
The quality of our results is somewhat limited by the use of the
``thinning" procedure, imposed by memory limitations of the
APE 6-Gflops machine \cite{APESEM}.
Our tentative conclusion is that the most appreciable improvement of the
chiral behaviour of the matrix element comes from the implementation of
the non-perturbative calculation of the mixing coefficients, whereas Clover
improvement and Boosted Perturbation Theory (BPT) seem to be less
influential. Preliminary results of this analysis have been presented
in \cite{DONINI}.
\par
Section \ref{sec:bk} introduces the  notation necessary for the
renormalization of the lattice operator;
 in section \ref{sec:details} we spell out all the details of the
numerical simulation; the main results of our work are contained in section
\ref{sec:chiral}; in particular the various fitting procedures are explained
and the effects of Clover improvement, boosting and non-perturbative
subtraction are analysed; in section \ref{sec:physic} we extract the
renormalization group invariant value of $B_K$;
 section \ref{sec:concl} contains our  conclusions.

\section{The renormalized operator}
\label{sec:bk}

The weak effective Hamiltonian relevant to $K^0$--$\bar K^0$ mixing is
\be
{\cal H}_{eff}^{\Delta S = 2} = C\left(M_W/\mu \right) \op (\mu)
\ee
where $\op(\mu)$ is the renormalized operator;
$C(M_W/\mu)$ is the corresponding Wilson coefficient in the
Operator Product Expansion (OPE) and
$\mu$ is the renormalization scale.
Our aim is to measure the  matrix element
$\langle {\bar K}^0 \vert {\hat O}^{\Delta S = 2}(\mu) \vert K^0 \rangle$,
parametrized by the $B_K$ parameter as follows:
\be
\label{kok}
\langle {\bar K}^0 \vert {\hat O}^{\Delta S = 2}(\mu) \vert K^0 \rangle =
\frac{8}{3} f^2_K m^2_K B_K(\mu) \, ,
\ee
where
\be \langle {\bar K}^0 \vert {\hat O}^{\Delta S = 2}(\mu) \vert K^0 \rangle \to
\frac{8}{3} \langle \bar K^0 | \bar s \gamma_{\mu} d | 0 \rangle
\langle 0 | \bar s \gamma_{\mu} d | K^0 \rangle  =\frac{8}{3} f^2_K m^2_K \ee
is the matrix element in the vacuum saturation approximation. \par
Since the lattice regularization with Wilson fermions explicitly breaks
chiral symmetry, the lattice $\op$ mixes under renormalization with
other ``effervescent"
operators with the same flavour numbers but with different
chirality \cite{BOCHICCHIO}--\cite{BORRELLI}:
\be
\label{contlatt}
{\hat O}^{\Delta S = 2}(\mu) = Z_+( \mu a , \alpha_s)
\left[ {\hat O}^{\Delta S = 2}\left( a \right)
+ Z_{SP} {\hat O}_{SP}\left( a\right) +
Z_{VA}{\hat O}_{VA}\left( a\right) +
Z_{SPT} {\hat O}_{SPT}\left( a\right) \right] \, ,
\ee
where $Z_{SP,VA,SPT}=Z_{SP,VA,SPT}(\alpha_s)$;
$\hat O(a)$ denotes the bare lattice operators
\bea
{\hat O}_{SP}\left( a\right)  &=&  - \frac{1}{16 N}
[\bar s d \bar s d - \bar s \gamma_5 d \bar s \gamma_5 d ] \, ;
\label{OSP} \\
{\hat O}_{VA}\left( a\right)  &=&  - \frac{N^2+N-1}{32 N}
[\bar s \gamma_{\mu} d \bar s \gamma_{\mu} d -
\bar s \gamma_{\mu} \gamma_5 d \bar s \gamma_{\mu} \gamma_5 d ] \, ;
\label{OVA} \\
{\hat O}_{SPT}\left( a\right)  &=&  \frac{N - 1}{16 N}
[\bar s d \bar s d + \bar s \gamma_5 d \bar s \gamma_5 d +
\bar s \sigma_{\mu\nu} d \bar s \sigma_{\mu\nu} d ] \, .
\label{OSPT}
\eea
and $N$ is the number of colours.
In the chiral limit, the
basis of operators appearing on the r.h.s. of eq. (\ref{contlatt})
remains unchanged  beyond one-loop, due to CPS symmetry \cite{BERNARD2}.
Note that, since there are no $\Delta S=2$ operators of lower dimension,
the mixing takes place between dimension-6 operators only.
This implies that the mixing coefficients are at most logarithmically
divergent and can be computed
on the lattice using Perturbation Theory (PT) \cite{BOCHICCHIO}.
This calculation has been performed to one-loop for both the Wilson and
Clover actions \cite{MARTIW,draper,BORRELLI}; the results are expressed in
terms of two constants $F_+$ and $F^*$ defined as
\be
\label{contlatt2}
Z_+=1 + \frac{\alpha_s}{4 \pi} F_+,\qquad
Z_{SP}=Z_{VA}=Z_{SPT}=\frac{\alpha_s}{4 \pi} F^*.
\ee
Notice that the equality between the off-diagonal
mixing coefficients holds only at one-loop.
Their values are given in table \ref{mix}.
In the Wilson case,
$F_+$ is the constant necessary to relate the bare lattice
operator to the continuum operator renormalized at $\mu=1/a$
 in the $\MSbar$  Dimensional Reduction scheme (DRED) \cite{MARTIW}.
The perturbative result for
$F_+$ in the Clover case requires some explanation.  In ref. \cite{BORRELLI},
$F_+=-10.9$ is the constant necessary to relate the bare lattice
operator to the continuum operator renormalized at $\mu=1/a$
 in DRED.
The value of $F_+$ reported in the table, instead, refers
to the overall lattice renormalization
constant necessary to normalize the operator
in the Regularization Independent (RI) scheme,
defined in ref. \cite{CIUCHINI2} and used in ref.
 \cite{TALEVI} in the Landau gauge ($F_+ \to F_+ -8 \ln 2 +5/3 + L(\mu)$).
This is the same scheme used in the non-perturbative case,
 to which we want to compare
the perturbative results.
In the RI scheme, there is a dependence of $F_+$ on
the momentum $p^2=\mu^2$ of the external states, which is
reflected in the term $L(\mu)=2/(16 \pi^2) \ln(\mu^2 a^2)$,
appearing in the table. This term
is numerically small in the range of $\mu^2 a^2$
used in our study: it gives a contribution of order
 $0.01$ to $Z_+$, if one uses the bare coupling,
and of order $0.02$ if one uses some boosted coupling,
see below;  for this reason, it will be ignored in the following.
\par
Although the renormalized operator $\op (\mu)$ is constructed so as to have
the correct chiral properties in the continuum limit, the perturbative
estimate of the $Z$'s is bound to introduce a systematic distorsion arising
from the one-loop truncation. BPT should improve
the behaviour of the perturbation series, thus providing better estimates of
the $Z$'s \cite{LEPAGE}. On the other hand, the  NPM
for the calculation  of the $Z$'s,
as proposed in ref. \cite{NP},
should in principle account for all perturbative and non-perturbative
contributions.
In order to compare the perturbative mixing coefficients to
the non-perturbative ones, in table \ref{mix} we also  give the
effective $F$'s
obtained  non-perturbatively in the Clover case  \cite{TALEVI}.
Since in the non-perturbative
case there is some  dependence on the renormalization scale $\mu^2 a^2$, see
also below, we give the results at two  values of $\mu^2 a^2$.
The $F$'s are defined from $Z_{SP,VA,SPT}
=\alpha^{(1)}_s/(4 \pi)
F_{SP,VA,SPT}$, with the $Z$'s  taken
from the non-perturbative calculation. We denote by
$\alpha^{(1)}_s=1.68\,  \alpha_s$
the coupling constant in the Boosted Scheme (BS) 1, to be
introduced in section \ref{sec:chiral},
and by $\alpha_s=
3/(2 \pi \beta )$  the bare lattice coupling constant.
\begin{table}
\centering
\begin{tabular}{|c|c|c|c|c|}\hline
Method  & $F_+$ & $F_{SP}$ & $F_{VA}$ & $F_{SPT}$ \\
\hline
\hline
W-PT & $-47.6$ & $9.6$  & $9.6$ & $9.6$ \\
SW-PT & $-14.8+ L(\mu)$  & $19.4$ & $19.4$ & $19.4$ \\ \hline
NPM-$\mu^2 a^2=0.96$ & $-15(3)$ & $13(7)$ & $28(2)$ & $23(4)$ \\
NPM-$\mu^2 a^2 =2.47$ & $-14(2)$ & $21(6)$ & $31(2)$ & $22(3) $ \\
\hline
\end{tabular}
\caption{\it{ Renormalization constants for the Wilson (W)
and SW-Clover (SW) actions.
The first two rows are the results from  one-loop perturbation theory,
denoted by $F_+$ and $F^*$ in the text.
The last two lines are the results for the effective constants
$F_{+,SP,VA,SPT}$, obtained from the corresponding
$Z_{+,SP,VA,SPT}$,
calculated  with the NPM in the Clover case [23],
 at two different  values of the renormalization
scale.}}
\label{mix} \end{table}

\par Were the coefficients of the mixing with the
dimension 6 operators to be known with infinite accuracy,
a further source of systematic error would still
be present, due to the  finiteness of the
lattice spacing.
The SW-Clover action \cite{SW} removes
$O(a)$ effects from all matrix elements, leaving us with $O(\alpha_s a)$
and $O(a^2)$
systematic corrections \cite{HEATLIE}.
It must be noted that in the Clover case all the results
refer to  the ``improved-improved" operators
introduced in refs. \cite{BORRELLI,MARTIR}.

\section{Computational details}
\label{sec:details}

The calculations have been done in the quenched approximation,
 generating  $460$ gauge configurations on a $18^3 \times 64$ lattice
at $\beta= 6.0$ with a standard Metropolis algorithm, using the 6-Gflops
version of APE.
The light fermion propagators have been computed on the
full set of configurations using the
SW-Clover action, and on a subset of 200 configurations using the Wilson
action.  The reason  for  this difference is the following.
Results obtained
with the two actions on the same ensemble of 200 configurations
 demonstrated that
the Wilson action is characterised by smaller statistical fluctuations.
In order to have comparable errors,  we then increased the statistics
to 460 configurations in the Clover case.
The values of the hopping parameter
used were $K = 0.1550, 0.1540, 0.1530$ for the Wilson  case and
$K$~$= 0.1440, 0.1432, 0.1425$ for the Clover case.
All two- and three-point
correlation functions have been calculated at degenerate quark masses.
 The statistical
error has been estimated by the jacknife method, by decimating 20 (46)
configurations at a time, for the ensembles of 200 (460) configurations.
\par
The $\hat O^{\Delta S=2}$ operator was placed at the origin.
In what follows,  $\op$ denotes the renormalized
operator defined in eq. (\ref{contlatt}).
One kaon source was  inserted at fixed time $t_y=12$ or $16$,
whereas the second kaon source   was allowed to vary over the
full  time-range  $t_x=0,63$;
each kaon source carries a  spatial momentum denoted by
$\vec p$ or  $\vec q$. In lattice units of $2 \pi/ L $ ($L$ is the
number of sites in each spatial direction) the values
considered were $\vec p_0=(0,0,0),\vec p_{\pm}=(\pm 1,0,0),
\vec p_{11}=(1,1,0)$ and the corresponding ones equivalent under
the cubic group.
The same values have also been attributed to $\vec q$.
The combinations of momenta examined were
$ (\vec p_0 , \vec q_0) $; $ (\vec p_0 , \vec q_{\pm}) $;
$ (\vec p_0 , \vec q_{11}) $; $ (\vec p_{\pm} , \vec q_{\pm}) $.
The last case
includes three possibilities of relative momentum orientation, namely parallel,
antiparallel and perpendicular. Cases which are equivalent under
the cubic group have been averaged.
\par
The meson two-point functions were fitted in the time interval $t_x=10$--$22$,
which guarantees a good isolation of the lightest pseudoscalar state.
By extrapolating (li\-near\-ly in the quark mass)
to the vanishing pseudoscalar mass, we computed  the critical
hopping parameter $K_{cr} = 0.15697(3)$  for the Wilson  action and
$K_{cr} = 0.14546(2)$ for the SW-Clover  action. The lattice spacing
determined from the $\rho$ mass is
$a^{-1} = 2.19(4)$ GeV (W) and $a^{-1} = 2.06(4)$ GeV (SW). Using the kaon
mass and interpolating the lattice pseudoscalar masses we find the hopping
parameter value at the strange quark mass, $K_s = 0.15468(9)$ (W) and
$K_s = 0.14366(7)$ (SW).
\par
In order to extract $B_K$ from
the  matrix element
$\langle \bar K^0(\vec p)|\hat O^{\Delta S=2}(\mu)|K^0(\vec q )\rangle$,
using the method exposed in section \ref{sec:chiral},
we calculated  the following
two- and three-point correlation functions
\bea
\label{asymp}
G_5(t_x;\vec p) & \equiv & \sum_{\vec x}\langle \bar s(x)
 \gamma_5 d(x) \bar d(0)
\gamma_5 s(0)\rangle e^{- i\vec p \cdot \vec x}\, , \nonumber \\
G_A(t_x) & \equiv & \sum_{\vec x}\langle \bar s(x) \gamma_0\gamma_5 d(x)
\bar d(0) \gamma_5 s(0)\rangle \, ,\\
G_{\hat O}(t_x,t_y;\vec p, \vec q) & \equiv & \sum_{\vec x,\vec y}
\langle \bar d(y) \gamma_5 s(y) \hat O^{\Delta S=2}(0) \bar d(x) \gamma_5 s
(x)\rangle
e^{-i\vec p \vec y} e^{i\vec q \vec x} \, ,\nonumber
\eea
and the ratios
\bea
\label{ratio}
R_3 & \equiv & \frac{G_{\hat O} (t_x,t_y;\vec p, \vec q)}
{G_5(t_x;\vec p) G_5(t_y;\vec q)}
\rightarrow \frac{\langle {\bar K}^0 (\vec p)
\vert {\hat O}^{\Delta S = 2} \vert K^0 (\vec q)\rangle}
{Z_5} \, , \\
X & \equiv & \frac{8}{3} \frac{G^2_A(t_x)}{G^2_5(t_x ;\vec 0)}
 \rightarrow
\frac{8}{3} \frac{f_K^2}{Z_A^2 Z_5} m_K^2 \, , \nonumber
\eea
where $Z_5 \equiv \langle 0|\bar s(0) \gamma_5 d(0)|K^0
\rangle$; $f_K$ is the kaon
decay constant and $Z_A$ is the axial current renormalization constant.
\begin{figure}   
    \begin{center}
       \setlength{\unitlength}{1truecm}
       \begin{picture}(6.0,6.0)
          \put(-4.,-5.){\includegraphics{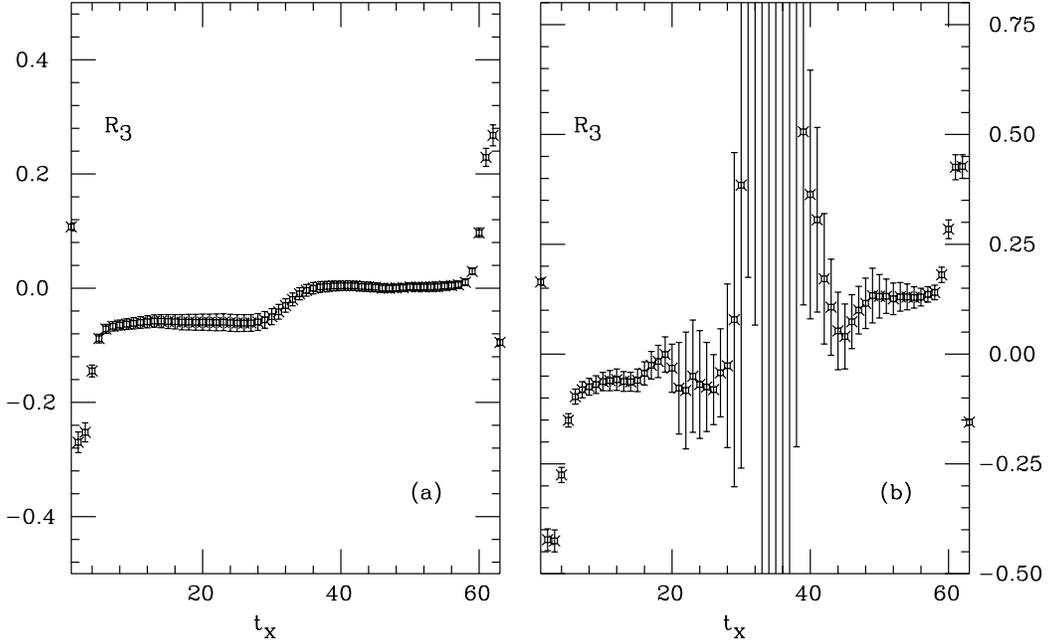}}
       \end{picture}
    \end{center}
\caption{\it{$R_3$ for the SW-Clover action at $K = 0.1440$
(a) $\vec p = \vec q = \vec p_0$; (b) average of three cases:
$\vec p = \vec q = \vec  p_+ = (1,0,0) $; $\vec p = \vec q = (0,1,0) $ ;
$\vec p = \vec q = (0,0,1) $.  The renormalized
operators have been obtained in  perturbation theory,
by using $\alpha^{(1)}_s$ as the boosted coupling, cf. eq. (17).}}
\label{plat}
\end{figure}
\par
The asymptotic behaviour indicated by the arrows
in the above equations is valid in the limit
$t_y \ll 0 \ll t_x$.
In the absence of final state interaction \cite{MAIANI}
and finite volume effects,
the ratio $R_3$ with  time ordering $0 \ll (t_y,t_x)$
  would isolate the matrix element $\langle \bar K^0(\vec p)
\bar K^0(-\vec q)|\hat
O^{\Delta S = 2}|0\rangle$.
 On our lattice with periodic boundary conditions
both signals appear as two distinct plateaux.
In fig. \ref{plat}, we show two typical cases
among the best (fig. \ref{plat}a) and
the worst (fig. \ref{plat}b) plateaux for $R_3$.
Two plateaux are clearly visible; the rightmost
one corresponds to the case of interest $\langle \bar K^0|\op|K^0\rangle$,
whereas the
leftmost one corresponds to $\langle \bar K^0 \bar K^0|\op|0\rangle$.
\par
Since one can  average those cases characterised by
spatial momenta equivalent under the cubic symmetry, we are faced with two
options: either average the correlation functions before taking their ratios
or average the ratios. Having found no significant difference between
the results obtained from the two methods, we have opted for the first one.
For example, in fig. \ref{plat}b, we show the ratio of the average of three
correlations.
\par
We  note that the results obtained from the data with $t_y = 16$ are
fully compatible to those obtained with $t_y = 12$; we only report results
from the latter case.  In order to extract the matrix elements and errors,
we have performed a  weighted   average of $R_3$ and $X$
over  $t_x$.
 The intervals in $t_x$ were chosen on the basis of  a study
of the time-dependence of the effective energy  (mass)
obtained from  $G_5$ for a
meson at rest.    In order to
have a good isolation of  the lightest pseudoscalar state, and avoid
exceedingly large statistical errors,
the optimal time interval was found to be $10 \le t_x \le 22$. This
was  the case
for both the Wilson and SW-Clover actions and for all the values of the hopping
parameter.
We estimate that the contamination from the excited states
induces a systematic error of some per cent  on  the relevant quantities
 for
all values of the quark masses considered in this study.
Since as meson sources  we used local operators,  whose  couplings
to the physical states are independent of the momentum,
 the same interval in $t_x$ has  been used for all
meson momenta.  Due to periodic boundary conditions, we also used
 the interval $42 \le t_x \le 54$. For
$R_3$, this corresponds to a different matrix element.
In the following, $X$ and $R_3$ will always denote the weighted averages
on $t_x$.
As mentioned above, computer memory limitations have obliged
us to use the so-called ``thinning'' approximation. Thinning has been
implemented here exactly as in \cite{APESEM} (which may be consulted
for details). Although we have explicitly tested that thinning does not
influence $G_5$, we have no way of testing its effect on more complicated
correlations, such as $G_{\hat O}$. We strongly suspect that this approximation
results in an appreciable worsening of the statistical quality of our data.
The rather large errors are also due to the relatively  small size of
our lattice.

\section{Chiral behaviour of the $B_K$ parameter}
\label{sec:chiral}

We now examine the chiral behaviour of the matrix element,
which we parametrize as
\bea
\label{koklatt}
\langle {\bar K}^0 \vert {\hat O}^{\Delta S = 2} \vert K^0 \rangle  =
\alpha + \beta m^2_K &+& \gamma ( p \cdot q ) + \nonumber \\
+ \delta m^4_K +  \epsilon m^2_K (p \cdot q) &+& \zeta (p \cdot q)^2 +
 \dots
\eea
where all the quantities are expressed in lattice units
and the ellipses stand for higher-order terms in
$(p \cdot q)$ and $ m^2_K$.
The matrix element on the l.h.s. of eq.~(\ref{koklatt}) is obtained by
multiplying $R_3$ of eq. (\ref{ratio}) by $Z_5$, which is extracted
in the standard way from the
asymptotic behaviour of the pseudoscalar two-point correlation $G_5$ of
eq. (\ref{asymp}).
 The parameters $\alpha$ and $\beta$ are
lattice artefacts that should vanish in the continuum limit. As we have already
stressed in section \ref{sec:bk}, their origin may
be attributed to two effects: (i) discretization errors of the lattice matrix
elements arising from the subtraction of operators with the wrong ``naive''
chirality; (ii) higher-order
perturbative corrections to the renormalization constants
$Z_+, Z_{SP}, Z_{VA}, Z_{SPT}$ of eq.~(\ref{contlatt}). Similar artefacts also
contaminate the other parameters $\gamma, \delta, \epsilon,\zeta$.
We investigate the differences in the values of
 $\alpha$ and $\beta$, obtained by fitting the dependence of the
matrix element on the kaon masses and momenta.
Since  in the continuum
$\alpha$ and $\beta$ vanish \cite{cabigel},  we consider a
reduction of their values  as a measure of the correctness of the chiral
behaviour and of the accuracy in the determination of the matrix element.
\par
There are several possibilities for extracting the parameters $\alpha,\beta,
\dots,\zeta$. We have verified that they all give compatible results.
In agreement with the observation of ref.~\cite{GAVELA}, we found that,
in order to minimize higher-order terms in $m_K^2$, the most convenient
procedure is to substitute the fitting variables as follows
\bea
m^2_K &\Rightarrow& X  =\frac{8}{3} \frac{f^2_K}{Z_5 Z^2_A} m_K^2\\
(p \cdot q) &\Rightarrow& Y =  \frac{X (p \cdot q)}{m_K^2}
 =  \frac{8}{3} \frac{f^2_K}{Z_5 Z^2_A} (p \cdot q)
\eea
and fit directly the ratio $R_3$ with respect to $X$ and $Y$.
Up to quadratic terms, with $\alpha, \beta, \dots, \zeta$ appropriately
redefined, eq. (\ref{koklatt}) becomes
\be
\label{koklatt2}
R_3 = \frac{1}{ Z_5 }
\langle {\bar K}^0 \vert {\hat O}^{\Delta S = 2} \vert K^0 \rangle
=  \alpha + \beta X + \gamma Y + \delta X^2  +  \epsilon X Y + \zeta Y^2
\, .\ee
Another advantage  is that $R_3$  (and $X$) are obtained directly
from  ratios of Green
functions which have correlated fluctuations. This has to be
contrasted with $\langle {\bar K}^0 \vert {\hat O}^{\Delta S = 2} \vert K^0
\rangle$ which is obtained by multiplying $R_3$ by $Z_5$, computed
from a fit of $G_5(t_x ; \vec 0)$.
The reduction of the size of the  higher-order terms in $m_K^2$,
obtained by using  $X$ and $Y$,  is reflected in the stability
of the results
 for $\alpha,\beta$ and $\gamma$. The values
obtained using eq. (\ref{koklatt2})
 differ by only about $20\%$  from
those obtained from the linear fit
\begin{equation}
\label{koklin}
R_3 = \frac{1}{Z_5} \langle {\bar K}^0 \vert {\hat O}^{\Delta S = 2}
\vert K^0 \rangle = \alpha + \beta X + \gamma Y \, .
\end{equation}
Using   eq. (\ref{koklatt}) instead, the results, although compatible
within the errors, would differ by  about $50 \%$.
Even though eq. (\ref{koklatt2}) more than halves the discrepancy between
linear and quadratic fits, the remaining systematic effect due to
higher-order corrections requires further
investigation.
Below, we only present results from eqs. (\ref{koklin}).
The $B$-parameter is computed  from the relation
\be B_K(\mu)=\gamma/Z_A^2\, . \label{gzag} \ee
\par
In the remainder of this section,
we discuss possible systematic effects on the  chiral behaviour
of the matrix element of $\op$.
\subsection{Chiral behaviour in perturbation theory}
One possible source of systematic error
 is that the one-loop perturbative determination
of the renormalization constants, at current values of $\beta$,
is not accurate enough to correct the  chiral behaviour.
Thus, it is essential to improve the accuracy of the mixing renormalization
constants which are calculated in PT. For this reason,
the by now standard procedure of
BPT \cite{LEPAGE} has been adopted for both actions;
we have tried two  different boosted schemes, BS1 and BS2,
by using the
following definitions of $\alpha_s$ \cite{LEPAGE}:
\begin{eqnarray}
\mbox{BS} 1: \qquad  \alpha^{(1)}_s &=&
\frac{1}{\langle \tr \Box \rangle} \alpha_s
\simeq 1.68 \,\alpha_s  \ \hbox{(W)} ; \ 1.68\,\alpha_s  \ \hbox{(SW)}
\\
\mbox{BS}2: \qquad  \alpha^{(2)}_s &=& (8 K_c)^4 \alpha_s \simeq 2.49
\, \alpha_s  \
\hbox{(W)};
\ 1.84 \, \alpha_s \ \hbox{(SW)}
 .\end{eqnarray}
As can be seen, the effective coupling depends quite significantly
on the prescription in the Wilson case, while the
variation in the Clover  case is small.
 In the following, we will also
give results with  $\alpha^{(3)}_s= 3.1 \, \alpha_s$ (BS3)
\cite{marbiele}, in the Wilson
case, in order to compare with \cite{BERNARD} and
will  call SPT the case in which we use the bare lattice
 $\alpha_s$ as expansion parameter.
\begin{figure}   
    \begin{center}
       \setlength{\unitlength}{1truecm}
       \begin{picture}(6.0,6.0)
          \put(-4.,-5.){\includegraphics{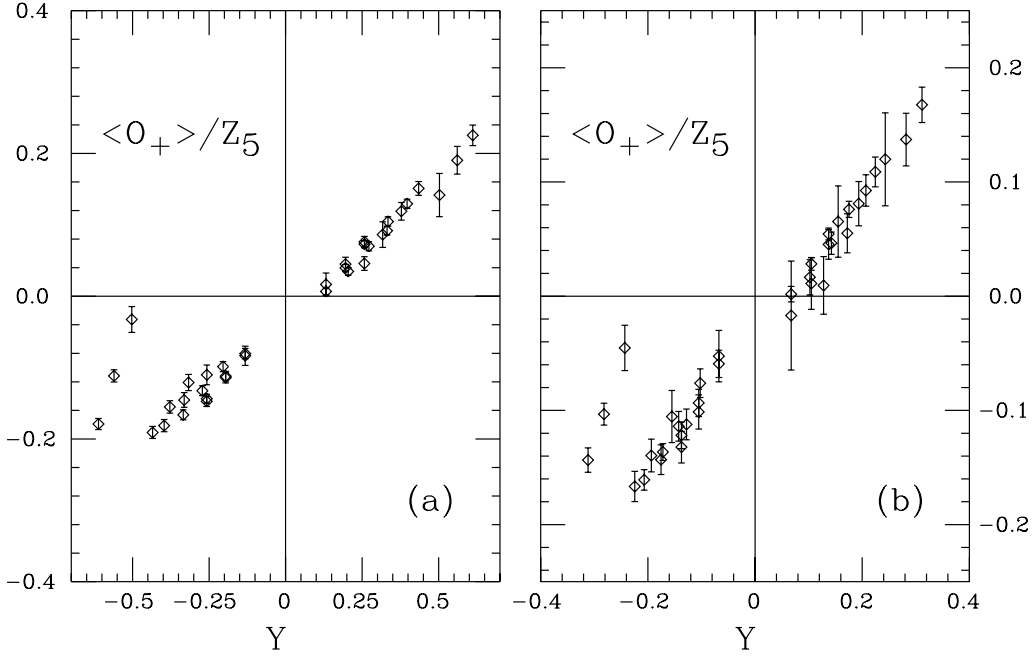}}
       \end{picture}
    \end{center}
\caption{\it{Dependence of
$\langle 0 \vert {\hat O}^{\Delta S = 2} \vert K^0 K^0 \rangle$ ($Y<0$) and
$\langle {\bar K}^0 \vert {\hat O}^{\Delta S = 2} \vert K^0 \rangle$
($Y>0$)
on mass and momentum obtained with (a) the Wilson action (200 configurations)
 and (b) the SW-Clover action (460 configurations). The renormalized
operators have been obtained by using $\alpha^{(1)}_s$ as
the boosted coupling.}}
\label{point}
\end{figure}
In fig. \ref{point},  we show all our results
for $R_3$, plotted against $Y= X (p\cdot q))/m_K^2$, where
$X=(8 f_K^2 m_K^2)/(3 Z_5 Z_A^2)$.
$Y > 0$ corresponds to the matrix element
$\langle \bar K^0|\op|K^0\rangle$,  whereas $Y < 0$
corresponds to $\langle \bar K^0 \bar K^0|\op|0\rangle$
(with the usual caveat of final state interaction \cite{MAIANI} and finite
size effects) \footnote{ Lattice artefacts are
clearly visible in the points corresponding to  the highest momenta
on the leftmost side of the figure.}.
Notice that the operator is renormalized in the $\overline{{\rm MS}}$
DRED scheme in the Wilson case, and in the RI scheme in the Clover case.
\par
In the following, we will only analyse  data
related to  $ \langle \bar K^0 \vert {\hat O}^{\Delta S = 2}
\vert K^0 \rangle$.
In  fig. \ref{point}, the effective coupling of the
BS1 (which  is the same for  the two quark actions)
was used to define the renormalized operator.
\begin{table}
\centering
\begin{tabular}{|c|c|c|c|}\hline
\rm{Renormalization} & $\alpha$ & $\beta$
 & $\gamma$  \\
\hline
\hline
\rm{SPT-W}    & $-0.100(7) $ & $0.21(8) $ & $ 0.56(7)$
  \\
\rm{BS1-W}  & $-0.068(5)$ & $0.15(5) $ & $ 0.40(5)$   \\
\rm{BS2-W}  & $-0.032(2)$ & $0.08(3) $ & $ 0.21(3)$   \\
\rm{BS3-W}  & $-0.0082(6)$ & $0.020(7) $ & $ 0.057(7)$   \\
\hline
\rm{SPT-SW}   & $-0.067(12) $ & $0.17(15)  $ & $ 0.62(11)$  \\
\rm{BS1-SW}         & $-0.054(12) $ & $0.17(15)  $ & $ 0.62(11)$  \\
\rm{BS2-SW}         & $-0.052(12) $ & $0.16(15)  $ & $ 0.62(11)$  \\
\hline
\rm{NPM-SW} $\mu^2 a^2 = 0.96$ & $0.017(13)$ &$0.21(17)$ & $ 0.70(12)$ \\
\rm{NPM-SW} $\mu^2 a^2 = 2.47$ & $0.022(14)$ &$0.23(17)$ & $ 0.72(12)$  \\
\hline
\end{tabular}
\caption{\it{SW-Clover (SW)  and Wilson (W) results with different choices of
mixing renormalization constants. For comparison, in the Clover case
the non-perturbative results at two values of the renormalization
scale are also given.}}
\label{result}
\end{table}
In table \ref{result}, we give the values of
 $\alpha$, $\beta$
  and $\gamma$ obtained from a fit of our data to eq. (\ref{koklin}).
Figure \ref{point} and the values of $\alpha$, $\beta$
  and $\gamma$
in table \ref{result} allow several observations:
\begin{enumerate}
\item  The use of the SW-Clover action
by itself does not have a significant effect on the chiral behaviour of
the matrix element.
\item For all the choices of boosted coupling,
the fitted values of $\alpha$ in table \ref{result}
are  different from  zero.
\item A comparison of  boosted and  unboosted results shows  that,
in the Clover case, $\alpha$, $\beta$ and $\gamma$ are
almost unaffected by the boosting procedure. This is to be contrasted
by the results in the
Wilson case, which are  rather unstable and strongly dependent on
the precise choice of the  effective coupling constant\footnote
{ Notice that in the literature boosting factors varying between
$1.6$ and $3.1$ have been used for this case, at $\beta=6.0$.}.
This is particularly relevant for $\gamma$, from which we
 estimate  the $B_K$ parameter. It implies that the estimate
of $B_K$ in the Wilson case is subject to a large systematic uncertainty,
due to its sensitivity to the choice of the expansion parameter,
i.e. to higher-order effects.
\item  In the Wilson case, the instability is due
to the large value of $F_+$
(see table \ref{mix}): $\alpha$, $\beta$ and $\gamma$
diminish rapidly with increasing $\alpha_s$, roughly
proportionally to  $Z_+$.

Inspired by mean field theory \cite{LEPAGE},
 we have used eq. (\ref{contlatt}), with $Z_+$ as an overall factor,
also in the perturbative case. Alternatively, we can   use the
formula
\be
\label{contlattp}
{\hat O}^{\Delta S = 2}(\mu) = Z_+( \mu a , \alpha_s)
 {\hat O}^{\Delta S = 2}\left( a \right)
+ Z_{SP} {\hat O}_{SP}\left( a\right) +
Z_{VA}{\hat O}_{VA}\left( a\right) +
Z_{SPT} {\hat O}_{SPT}\left( a\right)  \, ,
\ee
which is equivalent at one loop, but may lead to significant differences
for the matrix elements at  large values of $\alpha_s$.
Indeed, in the BS3 the two possibilities correspond to values of $\gamma$
which differ by  a factor of 2.5.
\end{enumerate}
\subsection{Chiral behaviour with the non-perturbative renormalization}
\begin{figure}   
    \begin{center}
       \setlength{\unitlength}{1truecm}
       \begin{picture}(6.0,6.0)
          \put(-4.,-5.){\includegraphics{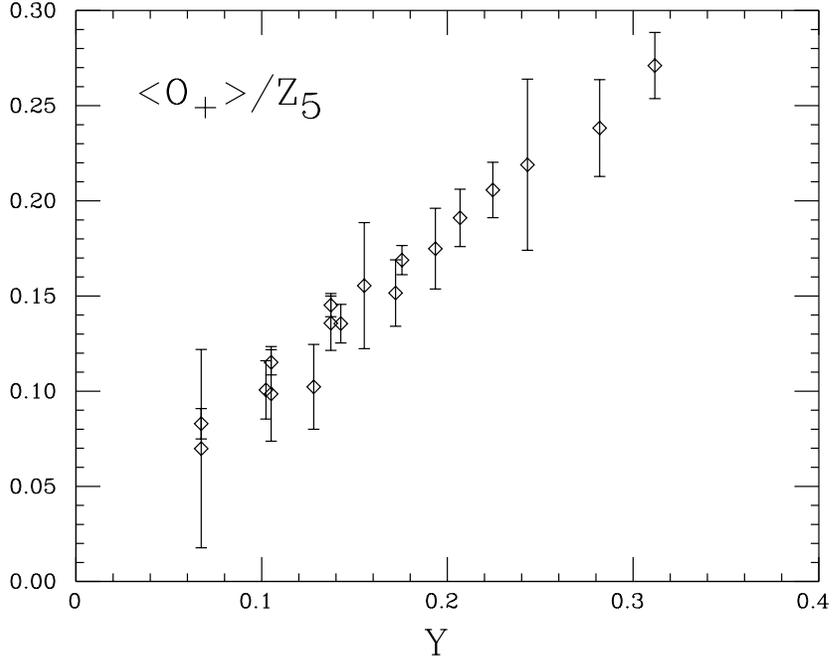}}
       \end{picture}
    \end{center}
\caption{\it{Dependence of
$\langle {\bar K}^0 \vert {\hat O}^{\Delta S = 2} \vert K^0 \rangle$
on mass and momentum, with the non-perturbative
renormalization constants computed at $\mu^2 a^2 =0.96$.
The data  can be compared with the corresponding perturbative ones
 in fig. 2.}}
\label{nper}
\end{figure}

We now examine the case in which  the renormalization constants are determined
non-perturbatively. This will only be done in the Clover
case, where the mixing coefficients have  already been
computed. The data for the operator
renormalized at $\mu^2 a^2=0.96$ are shown in fig. \ref{nper}.
 In ref. \cite{TALEVI},  the non-perturbative
method  of ref. \cite{NP} was
 implemented for  the calculation of $Z_+, Z_{SP}, Z_{VA}$ and $Z_{SPT}$.
For completeness, we reproduce the results of ref. \cite{TALEVI}
in table \ref{tab:examples}.
\begin{table}
\centering
\begin{tabular}{|c|c|c|c|c|}
\hline
$\mu^2 a^2$ &$Z_+$ & $Z_{SP}$ & $Z_{VA}$& $Z_{SPT}$  \\
\hline \hline
$0.46$ & $0.91 \pm 0.05$ & $0.08 \pm 0.14$ & $0.34 \pm 0.03$ & $0.34 \pm 0.07$
      \\
$0.66$ & $0.84 \pm 0.05$ & $0.07 \pm 0.09$ & $0.33 \pm 0.03$ & $0.34 \pm 0.06$
      \\
$0.81$ & $0.83 \pm 0.04$ & $0.11 \pm 0.07$ & $0.31 \pm 0.03$ & $0.28 \pm 0.05$
      \\
$0.96$ & $0.84 \pm 0.03$ & $0.14 \pm 0.07$ & $0.30 \pm 0.02$ & $0.24 \pm 0.04$
       \\
$1.27$ & $0.80 \pm 0.04$ & $0.17 \pm 0.05$ & $0.29 \pm 0.02$ & $0.21 \pm 0.03$
      \\
$1.54$ & $0.82 \pm 0.02$ & $0.19 \pm 0.04$ & $0.27 \pm 0.02$ & $0.16 \pm 0.03$
       \\
$1.89$ & $0.83 \pm 0.03$ & $0.22 \pm 0.05$ & $0.30 \pm 0.02$ & $0.18 \pm 0.03$
       \\
$2.47$ & $0.85 \pm 0.02$ & $0.22 \pm 0.06$ & $0.33 \pm 0.02$ & $0.23 \pm 0.03$
        \\ \hline
SPT    & $0.91$ & $0.12$ & $0.12$ & $0.12$
       \\
BPT    & $0.84$ & $0.21$ & $0.21$ & $0.21$
      \\
\hline
\end{tabular}
\caption{\it
{Values of the  $Z$'s  for several
  renormalization scales $\mu^2 a^2$. We also give the results
obtained at $\mu^2 a^2=1$, by using ``standard"  perturbation theory
(SPT) and ``boosted" perturbation theory (BPT) with the BS1 effective
coupling $\alpha_s^{(1)}$.}}
\label{tab:examples}
\end{table}
With the NPM, the renormalization conditions for $\op$ can be
 imposed at  any arbitrary  scale $\mu^2  a^2$,
subject to the conditions that lattice artefacts
are small and that $\mu \gg \Lambda_{{\rm QCD}}$, so that
continuum perturbation theory can be applied.
In \cite{TALEVI},
a window of values around $\mu^2 a^2 \simeq 1$ was
established, for which the non-perturbative estimates of the $Z$'s
are fairly stable, see table   \ref{tab:examples};
the results in table \ref{tab:parameters}
show that, with the non-perturbative renormalization,
 $\alpha$ and $\beta$ are always compatible with zero  and
 $\gamma$ is very stable (for $\mu^2 a^2 \ge 0.66$)
in all the range of scales considered in our study.

The use of the  values of the $Z$'s in table \ref{tab:examples}
is subject at present to some limitations that we briefly want to
discuss\footnote{ These limitations are  not intrinsic to the method and
 will be easily eliminated in  future calculations.}.
\begin{enumerate} \item The calculation of ref. \cite{TALEVI}
was   performed on 36 configurations
at $\beta = 6.0$ on a $16^3 \times 32$ lattice, using   the SW-Clover action
and ``improved-improved" operators. The renormalization constants
were determined  for a single quark mass only,
corresponding to $ K = 0.1425 $.
We used these constants at all values of the quark
masses. Even though for light quarks
we expect very little dependence of the $Z$'s on the quark mass
\cite{ZETA_V,MARTIR},  one has to bear in mind that this approximation
may introduce a small systematic effect in our results.
\item The optimal strategy to obtain an accurate determination
of the matrix element
of the renormalized operator would be to compute the renormalization constants
and the operator matrix elements on the same set of configurations and
for the same values of the quark masses.
This is not possible  at the present stage,
 since the $Z$'s and the matrix elements were computed on different
sets of configurations, quark masses  and lattice volumes. On the other
hand, estimating the uncertainty of the final result
by varying the $Z$'s and the matrix elements independently within their errors
is questionable because the values of the $Z$'s, obtained by solving
a linear system of equations on the same configurations, are strongly
correlated. For this reason,
 we used the central values of
 the $Z$'s  from \cite{TALEVI} and  ignored
their  statistical error in the present analysis.
To monitor the stability of the results, we varied over a wide range
the renormalization scale
at which the $Z$'s were computed. Correspondingly, the
 $Z$'s change by an amount  comparable  to their error, or even larger.
A large set of results, obtained from a fit of the
data to eq. (\ref{koklin}),  corresponding
to $0.19 \le \mu^2 a^2 \le 2.47$ is given in table \ref{tab:parameters}.
Notice that for scales below $0.81$, the values of the $Z$'s
become unstable and
the use of perturbation theory (for matching) questionable; for
$\mu^2 a^2 \ge 2.47$ lattice artefacts become clearly visible
\cite{TALEVI,NP}.
\begin{table}
\centering
\begin{tabular}{|c|c|c|c|c|}
\hline
$\mu^2 a^2$ & $\alpha$ & $\beta$& $\gamma$ \\ \hline \hline
$0.19$ & $ 0.03(2)   $ & $0.2(2)   $ & $ 0.99(17) $\\
$0.46$ & $ 0.022(16) $ & $0.23(19) $ & $ 0.78(13) $\\
$0.66$ & $ 0.013(15) $ & $0.20(17) $ & $ 0.70(12) $\\
$0.81$ & $ 0.012(13) $ & $0.21(17) $ & $ 0.69(12) $\\
$0.96$ & $ 0.017(13) $ & $0.21(17) $ & $ 0.70(12) $\\
$1.27$ & $ 0.015(13) $ & $0.21(16) $ & $ 0.66(11) $\\
$1.54$ & $ 0.018(13) $ & $0.22(16) $ & $ 0.67(12) $\\
$1.89$ & $ 0.023(13) $ & $0.22(16) $ & $ 0.69(12) $\\
$2.47$ & $ 0.022(14) $ & $0.23(17) $ & $ 0.72(12) $\\
\hline
\end{tabular}
\caption{\it{ Values of the parameters of the fit of
$\langle \bar K^0 \vert {\hat O}^{\Delta S = 2} \vert K^0 \rangle$
to eq. (15).}}
\label{tab:parameters}
\end{table}
\item The stability of the results for the matrix elements of the
renormalized operator appears  surprising, since the renormalization
constants vary appreciably on  the same range of $\mu^2 a^2$
\cite{TALEVI}.
The explanation for this is the following.
The most unstable of the renormalization constants, according to
ref. \cite{TALEVI}, is $Z_{SP}$. The
$\hat O_{SP}$ matrix element carries, however, the least weight
in the final result. On the other hand, the most stable of
the $Z$'s is $Z_{VA}$, which gives the largest contribution.
The case of the operator $\hat O_{SPT}$ is intermediate between the two
cases discussed before.  We show this explicitly in
table \ref{separ}, where we present  separately, for different quark masses,
 the contribution of
the different operators to the final result.
\begin{table}
\centering
\begin{tabular}{|c|c|c|c|c|c|}\hline
Method&  $K$ & $R^{\Delta S = 2}$ & $R_{SP}$ & $R_{VA}$
& $R_{SPT}$  \\
\hline
Wilson & 0.1550    & $-0.004(5)$ & $-0.0154(4)$ & $0.0575(13)$ & $-0.0313(7)$
\\
BS1    & 0.1540    & $ 0.027(4)$ & $-0.0161(3)$ & $0.0620(10)$ & $-0.0331(5)$
\\
$$     & 0.1530    & $ 0.058(4)$ & $-0.0165(2)$ & $0.0654(8)$ & $-0.0345(4)$ \\
\hline \hline
Clover & 0.1440  & $-0.024(5)$ & $-0.0491(8)$ & $0.178(3)$ & $-0.1026(15)$ \\
BS1    & 0.1432  & $-0.002(5)$ & $-0.0515(7)$ & $0.190(3)$ & $-0.1083(13)$ \\
$$     & 0.1425  & $ 0.020(4)$ & $-0.0529(6)$ & $0.199(3)$ & $-0.1123(12)$ \\
\hline
NPM-Clover & 0.1440
   & $-0.024(5)$ & $-0.0332(6)$ & $0.261(5)$ & $-0.1202(17)$ \\
$\mu^2 a^2=0.96$ &0.1432
 & $-0.002(5)$ & $-0.0348(5)$ & $0.279(4)$ & $-0.1315(14)$ \\
$$ & 0.1425    & $ 0.020(4)$ & $-0.0358(4)$ & $0.292(4)$ & $-0.1315(14)$ \\
\hline
NPM-Clover& 0.1440 & $-0.025(5)$ & $-0.0540(9)$ & $0.284(5)$ & $-0.1161(17)$ \\
$\mu^2 a^2=2.47$&0.1432 &
 $-0.002(5)$ & $-0.0567(8)$ & $0.305(5)$ & $-0.1225(15)$ \\
$$& 0.1425 & $ 0.021(4)$ & $-0.0582(7)$ & $0.319(4)$ & $-0.1270(14)$ \\
\hline
\end{tabular}
\caption{\it{ Separate contributions
 to $R_3$ from the bare operator $R^{\Delta S = 2}$
and the off-diagonal matrix elements, $R_{SP}$, $R_{VA}$
and  $R_{SPT}$, for different
 quark masses. We give the non-perturbative results at two
different renormalization scales, and the perturbative result in the BS1. All
the matrix elements correspond to the case where the two mesons are at rest.}}
\label{separ}
\end{table}
For comparison, we also give the different contributions in the perturbative
case in the BS1.
In the table,  the
contribution to $R_3$ of the bare operator $\op$, corresponding
to $ \langle \bar K^0 \vert {\hat O}^{\Delta S = 2}
\vert K^0 \rangle/Z_5$, will be denoted by $R^{\Delta S = 2}$;
the contribution  of the operator $\hat O_{SP}$,
corresponding to $ \langle \bar K^0 \vert \hat O_{SP}
\vert K^0 \rangle/Z_5$, etc., will be denoted by  $R_{SP}$, etc.
We see that the largest subtraction comes
from the operator $\hat O_{VA}$, which is multiplied by the
constant $Z_{VA}$, which is quite well determined.
This explains the stability of the fitting parameters
$\alpha,\beta$ and $\gamma$  with varying renormalization scale.
\item The improvement in the chiral behaviour of the matrix element
arises from the difference in $Z_{SP}:Z_{VA}:Z_{SPT}$
between the non-perturbative and perturbative case.
Since  $Z_{VA}$ is much larger than $Z_{SP}$ and
$Z_{SPT}$  in  the  non-perturbative case,
 most of the beneficial effect on the chiral behaviour is due
to the enhanced contribution of $\hat O_{VA}$.
  Boosted perturbation theory, on the other hand,
 can only change  all the coefficients by the same amount,
leaving $Z_{SP}:Z_{VA}:Z_{SPT}$ unaltered, so that  the
correction is much less effective. A
very large boost of the coupling
 would be necessary
in order to obtain the same effect as in the non-perturbative case.
\end{enumerate}
\section{Physics results}
\label{sec:physic}
  In order to compare the results for
 the renormalized $B$-parameter obtained from $\gamma$,
cf. eq. (\ref{gzag}), with other  theoretical predictions of the same quantity,
it is useful to refer to a ``standard" definition, which we will take to be
the renormalization group
 invariant $B$-parameter $\hat B_K$.
In the standard perturbative approach, by a suitable choice of $Z_+$,
the $B$-parameter is defined
in the $\MSbar$ scheme at a scale $\mu \sim 1/a$. Let us call this
$B$-parameter
$B^{\overline{MS}}(\mu)$.
 At the next-to-leading order (NLO), $\hat B_K$ is written in terms
of  $B^{\overline{MS}}(\mu)$ as
\begin{equation}
\label{rgi}
{\hat B}_K = \alpha_s (\mu)^{-\gamma^{(0)}/2 \beta_0}
\left[ 1- \frac{\alpha_s (\mu)}{4\pi}
\left(\frac{\gamma^{(1)} \beta_0-\gamma^{(0)} \beta_1}{2\beta^2_0}\right)
\right] B^{\overline{MS}}(\mu)\, , \label{rgibp}
\end{equation}
where $\beta_{0,1}$ and $\gamma^{(0,1)}$ are
the leading and next-to-leading coefficients of the $\beta$-function and
anomalous dimension.  $\beta_{0,1}$ and $\gamma^{(0)}$ are universal
while $\gamma^{(1)}$ depends on the renormalization scheme.
The explicit expressions of  all the quantities appearing in
  eq. (\ref{rgibp}), $\gamma^{(0),(1)}$ and
$\beta_{0,1}$, can be found for example
in refs. \cite{BURAS,Ciuchini}. Notice that the renormalization group invariant
$B$-parameter defined in this way  is also regularization-scheme
independent, up to next-to-next-to-leading order terms.
\par We compute $\hat B_K$  only for the Clover case using the NPM,
since this is our best result.
By using  the non-perturbative
estimate of $Z_A = 1.06 $\cite{ZETA_A},
we obtain the $B$-parameter corresponding to the
operator renormalized in
 the  Regularization Independent (RI) scheme \cite{CIUCHINI2,NP}.
The matching to the $\MSbar$ scheme can be done in continuum perturbation
theory (this is one of the main advantages of the non-perturbative approach),
 by computing
the operator matrix element in the same gauge and on the same
external quark states as those used for the
 non-perturbative  calculation
of the lattice renormalization constants \cite{TALEVI,NP}.
At the next-to-leading order, the
 NDR--RI matching coefficient for $\op$  is given by \cite{CIUCHINI2}
\be B^{\overline{MS}}(\mu)=
\left(1+\frac{\alpha_s(\mu)}{4\pi}\Delta r_+^{\overline{MS}}\right)
B_K(\mu) \ee
where
\be
\Delta r_+^{\overline{MS}}=-14/3+8\log(2)\, . \ee
In numerical estimates, we have used
$\alpha_s(\mu)=0.3$, at  the
renormalization scale  $\mu^2 a^2 = 0.96$, corresponding to $\mu \sim 2$
GeV.  Given the still large
error on $B_K$, we have not included  the
uncertainty due to the choice of $\alpha_s(\mu)$ in the result given below.
At fixed $\alpha_s(\mu)$, there is  still a variation of $\pm 0.03$
when varying  $n_f$, the number of active flavours, between $0$ and $4$
in eq. (\ref{rgibp}).
With the above parameters, we get
\be \hat B_K= 0.86 \pm 0.15 \, .\ee

\par
Within large errors,
our results  agree with other lattice calculations
and with other results obtained with a different theoretical approach, namely
 the $1/N_c$ expansion \cite{PRADES,BARDEEN}.
 It may be noted that, in spite of our increased
statistics, the final errors for our results
 are comparable (or even larger) to those of previous
lattice studies. We attribute this to our thinning approximation, which we
plan to remove in  future calculations.

\section{Conclusions}
\label{sec:concl}
We have presented a parallel lattice calculation of the kaon
$B$ parameter $B_K$ with the Wilson  and the SW-Clover actions,
at $\beta=6.0$. In the Wilson case, using perturbation
theory to determine the renormalized four-fermion operator,
we found that the value of $B_K$ strongly depends on the precise
choice of the expansion parameter $\alpha_s$ and on higher-order
corrections. Consequently the results are subject
to a rather large uncertainty. In the Clover case, the
results were
more  stable with respect to a change of $\alpha_s$. The most appreciable
improvement in the chiral behaviour of the operator is,  however, found
with  the operator renormalized non-perturbatively,
by using the mixing coefficients computed  in  ref. \cite{TALEVI}.
\par
The accuracy of our  results was somehow limited by  two reasons:
the precision in the calculation of the mixing coefficients
and  the memory  of the
APE 6-Gflops machine, which forced us to work on  a relatively small
lattice volume and to use the ``thinning" approximation.
 These limitations are  not intrinsic to the non-perturbative  method and
 will be easily eliminated in  future calculations. In particular,
we expect  significant improvements
  by  computing  renormalization constants
and  operator matrix elements on the same set of configurations and
for the same values of the quark masses (and, of course,
by eliminating the ``thinning" and by working on a larger volume).
We also plan to compute non-perturbatively the mixing coefficients in   the
Wilson case, although, in this case,
 discretization errors could still play an important role.
In the non-perturbative case,
with  more accurate results, it  will be possible to check whether
$\alpha$ and $\beta$ are so  small as to give a negligible
contribution to the matrix element  for quark masses corresponding to the kaon.
Then, similarly to the case of staggered fermions,
 it will be possible to compute the kaon  matrix element
directly, without the  fitting procedures used so far.
This will allow us then to  reduce the error in the determination of
the quenched  $\hat B_K$  to about $5 \%$, as in the staggered case.
\par The results of our study are
encouraging, and motivate us
 to extend the calculation of  matrix
elements of operators renormalized  non-perturbatively
to the operators relevant to $\Delta I=1/2$
transitions and to the penguin and electro-penguin
operators which control CP-violation in kaon systems.
Our  strategy is   to achieve an
accurate determination of the physical weak amplitudes,
 by combining the improvement of the action
{\em \`a la} Symanzik, which reduces $O(a)$ effects, with the
non-perturbative method of ref. \cite{NP}.
\section*{Acknowledgements}
We warmly thank C.R. Allton
 for an early participation to this work,
 R.~Frezzotti, M. Guagnelli, G.~C.~Rossi,
C.T. Sachrajda and G. Salina  for many useful discussions
and  the members of the APE collaboration for their precious help.
M.T. thanks the Theory Division of CERN for the
 kind hospitality during the completion of this work.
We acknowledge the partial support by  M.U.R.S.T., Italy.

\end{document}